\documentclass[epj-spec]{svjour}
\usepackage{graphics}
\usepackage{amssymb,amsmath}
\usepackage{graphicx}
\usepackage{multicol}
\usepackage{multirow}
\usepackage{slashbox}
\usepackage{subfigure}
\usepackage{epsf}
\usepackage{hyperref}
\usepackage{cite}

\newcommand{\be}{\begin{equation}}
\newcommand{\ee}{\end{equation}}
\begin{document}
\title{Axial Anomaly and Mixing Parameters of Pseudoscalar Mesons}
\author{Yaroslav N. Klopot\inst{1}\fnmsep\thanks{\email{klopot@theor.jinr.ru}} \and Armen G. Oganesian\inst{2}\fnmsep\thanks{\email{armen@itep.ru}} \and Oleg V. Teryaev\inst{1}\fnmsep\thanks{\email{teryaev@theor.jinr.ru}} }
\institute{Joint Institute for Nuclear Research, Bogoliubov Laboratory of Theoretical Physics, Dubna 141980, Russia \and Institute of Theoretical and Experimental Physics, B.Cheremushkinskaya 25, Moscow 117218, Russia}
\abstract{
In this work the analysis of mixing parameters of the system involving $\eta$, $\eta'$
mesons and some third massive state $G$ is carried out. We use the generalized mixing scheme with three angles. The framework of the dispersive approach to Abelian
axial anomaly of isoscalar non-singlet current and
the analysis of experimental data of charmonium radiative decays
ratio allow us to get a number of quite strict constraints for the mixing
parameters. The analysis shows that the equal values of axial
current coupling constants $f_8$ and $f_0$ are preferable which may be considered
as a manifestation of $SU(3)$ and chiral symmetry.
} 
\maketitle

\section{Introduction} \label{intro}
This work is developing the approach of the papers \cite{Klopot:2008ec,Klopot:2009cm} and is devoted to the significant problem of mixing of pseudoscalar mesons. It is especially important with a number of current and planned experiments.

The problem of $\eta$-$\eta'$ mixing  has been studied for many years. The usual approach with one mixing angle dominated for decades, but in the recent years the more elaborated schemes appear to be unavoidable \cite{Feldmann:1998sh,Feldmann:1998vh,DeFazio:2000my,Kroll:2005sd,Escribano:2005qq,Mathieu:2009sg}. In particular, the theoretical ground of this  was based on the recent progress in the ChPT \cite{Leutwyler:1997yr,Kaiser:1998ds,Kaiser:2000gs}. On the other hand, it was shown, that the current experimental data cannot satisfactory describe the whole set of experiments within the one-angle mixing scheme.

The mixing schemes are usually enunciated either in terms of $SU(3)$ or quark basis. In our paper \cite{Klopot:2009cm} we construct and use the generalization of $SU(3)$ basis similar to the mixing of massive neutrinos. This is because we use the dispersive approach to axial anomaly (\cite{Dolgov:1971ri}, \cite{Ioffe:2006ww} for a review) to find some model-independent and precise restriction on the mixing parameters.

It was shown that any scheme with more than one angle unavoidably demands an additional  admixture of higher mass state. If we restrict ourselves to only one additional state $G$ (denoted as a glueball without really specifying its nature) then the general mixing scheme can be described in terms of 3 angles. In particular cases the number of angles can be reduced to two.

In the paper \cite{Klopot:2009cm} the analysis of different conventional (and most physically interesting) particular cases was performed (including two-angle mixing schemes) basing on the dispersive representation of axial anomaly from one side and charmonium decays ratio from the other side.

The main conclusion of the paper \cite{Klopot:2009cm} is that in all considered cases the only reasonable solutions appear at $f_8=f_0\simeq f_\pi$. The main aim of this work is to check whether this relation remains valid in the most general case with some specific constraints imposed.

This paper is organized as follows. In the Sec. \ref{Mixing} we introduce our notation and the general approach to the mixing. In Sec. \ref{Anomaly&R} we derive the basic equations relying on the dispersive approach to  Abelian axial anomaly of isoscalar non-singlet current $J_{\mu 5}^8$ and the  charmonium radiative decay ratio $R_{J/\Psi}$, while in Sec. \ref{Analysis} we perform the numerical analysis of these equations. Finally, in Sec. \ref{Conclusion} we present the conclusion.

\section{Mixing scheme} \label{Mixing}
We start with a ($N$-component) vector of physical pseudoscalar fields consisting
of the fields of the lightest pseudoscalar mesons and other fields:
\be \widetilde{\mathbf{\Phi}}\equiv
\begin{pmatrix}
\pi^0\\
\eta\\
\eta'\\
G \\
\vdots
\end{pmatrix}.
\ee

We are not able to specify the physical nature of the other
components with higher masses, the lowest of which $G$ can be
either a glueball or some excited state \footnote{Note, that the mixing with the excited states is usually(e.g. \cite{Escribano:2007nt,Cheng:2008ss}) supposed to be suppressed.}. Let us also introduce, following
\cite{Ioffe:1979rv,Ioffe:1980mx}, a set of $SU(3)$ fields
$\varphi_3, \varphi_8, \varphi_0$ ($\Phi_1,\Phi_2,\Phi_3$) and complement them with other (sterile) fields $g_i$ ($\Phi_i,i=4..N$)

\be
\label{Phi} \mathbf{\Phi}=
\begin{pmatrix}
\varphi_3\\
\varphi_8\\
\varphi_0\\
g\\
\vdots
\end{pmatrix}.\;\;\;
\ee

The three upper fields $\varphi_3, \varphi_8, \varphi_0$ are the only ones which define the generalized PCAC relation for axial current $J_{\mu5}^a=\overline{q}\gamma_\mu\gamma_5\frac{\lambda^a}{\sqrt{2}}q$ (no summation over $a$ contrary to $j$ and $k$ is assumed):

\be \label{J_diverg1} \partial_\mu J_{\mu5}^a =f_{a}\frac{\delta \Delta \mathcal{L}}{\delta \Phi_a} =F_{aj} M_{jk} \Phi_k, \;\; a=3,8,0,\;\; j,k=1..N,
\ee
where $\Delta \mathcal{L}$ is the mass term  in the effective Lagrangian with a non-diagonal mass matrix $\mathbf{M}$ (as fields $\Phi_k$ are not orthogonal to each other):

\be\Delta \mathcal{L}= \frac{1}{2} \mathbf{\Phi^T M \Phi},
\ee
and $\mathbf{F}$ is a matrix of decay constants \footnote{Note,
that matrix of decay constants $\mathbf{F}$ is  non-square
expressing the fact that generally the number of $SU(3)$ currents
is less then the number of all possible states involved in
mixing. The similar situation takes place (see e.g.
\cite{Bilenky:1998dt}) in one of the extensions of the Standard
Model -- neutrino mixing scenario involving sterile neutrinos.}:

\be
\mathbf{F}\equiv \begin{pmatrix}
f_3 & 0 & 0 & 0 & \ldots & 0\\
0 & f_8 & 0 & 0 & \ldots & 0\\
0 & 0 & f_0 & 0 & \ldots & 0\\
\end{pmatrix}.
\ee

In order to proceed from initial $SU(3)$ fields $\mathbf{\Phi}$ to physical mass fields $\widetilde{\mathbf{\Phi}}$ the unitary (real, as the CP-violating effects are negligible) matrix $\mathbf{U}$ is introduced

\be \mathbf{\widetilde{\Phi}=U\Phi} \ee that diagonalizes
the mass matrix \be \mathbf{UMU^{T}=\widetilde{M}}\equiv
diag(m_{\pi^0}^2,m_{\eta}^2,m_{\eta'}^2,m_G^2, \ldots), \ee
where $m_\pi$, $m_{\eta}$, $m_{\eta'}$ and $m_{G}$ are the masses
of the $\pi$, $\eta$, $\eta'$ mesons and glueball state $G$,
respectively.

Simple transformations of Eq.(\ref{J_diverg1}) read:
\be \label{divJ}
\mathbf{\partial_\mu J_{\mu5} =
FU^{T}\widetilde{M}\widetilde{\Phi}}
\ee
This formula is close to those obtained in
\cite{Ioffe:1979rv,Ioffe:1980mx} (in the limit of small mixing).
When the decay constants are equal, it is reduced to formula
(3.40) in \cite{Diakonov:1995qy}.

The matrix elements of $\mathbf{\partial_\mu J_{\mu5}}$ between vacuum state and physical states $\vert \widetilde{\Phi_k} \rangle$

\be \label{divJ-ME}
\langle 0 \vert \partial_\mu J_{\mu5}^a \vert \widetilde{\Phi_k} \rangle = F^a_i (U^{T}\widetilde{M})^i_k
\ee
can be compared to the standard definition of the "physical" coupling constants of axial currents:
\be \label{dec-const}
\langle 0 \vert J_{\mu5}^a \vert \widetilde{\Phi_k} \rangle= i f^a_k q_\mu.
\ee
From  (\ref{divJ-ME}) and (\ref{dec-const}) follows the relation
\be \label{Const}
f^a_k=F^a_i (U^{T})^i_k=f_a (U^T)^a_k.
\ee
This expression (recall, that there is no summation over $a$ ) clearly shows that $f^a_k$ are obtained by multiplication of each line of $\mathbf{U^T}$ by respective coupling $f_a$ and form a non-diagonal (contrary to $\mathbf{F}$) matrix.

Taking into account the well-known smallness of $\pi^0$ mixing
with the $\eta, \eta'$ sector
\cite{Ioffe:1979rv,Ioffe:1980mx,Ioffe:2007eg} and neglecting all
higher contributions we restrict our consideration to three
physical states $\eta, \eta', G$ and two currents $J_{\mu5}^8,
J_{\mu5}^0 $.  Then the divergencies of the axial currents
(recall, that $G$ is a first mass state heavier than $\eta'$):
\be \label{DivJ_FUMPhi}
\begin{pmatrix}
\partial_\mu J_{\mu5}^8\\
\partial_\mu J_{\mu5}^0\\
\end{pmatrix}
=
\begin{pmatrix}
f_8 & 0 & 0\\
0 & f_0 & 0\\
\end{pmatrix}
\mathbf{U^{T}}
\begin{pmatrix}
m_{\eta}^2 & 0 & 0\\
0 & m_{\eta'}^2 & 0\\
0 & 0 & m_{G}^2
\end{pmatrix}
\begin{pmatrix}
\eta\\
\eta'\\
G
\end{pmatrix}.
\ee

Exploring the mentioned similarity of the meson and lepton mixing,
we use the Euler parametrization for the mixing
matrix $\mathbf{U}$ (we use notation $c_i\equiv cos\theta_i, s_i\equiv
sin\theta_i$):

\begin{eqnarray}
\mathbf{U}=
\begin{pmatrix}
c_8c_3-c_0s_3s_8 & -c_3s_8-c_8c_0s_3 & s_3s_0\\
s_3c_8+c_3c_0s_8 & -s_3s_8+c_3c_8c_0 & -c_3s_0\\
s_8s_0 & c_8s_0 & c_0
\end{pmatrix}.  \label{MixMat}
\end{eqnarray}

In the following consideration we will need the divergency of the octet current $\partial_\mu J_{\mu5}^8$, so let us write it out explicitly:

\be \label{DivJ8}
 \partial_\mu J_{\mu5}^8=f_8(m_{\eta}^2 \eta (c_8c_3-c_0s_3s_8)+m_{\eta'}^2 \eta' (s_3c_8+c_3c_0s_8)+m_{G}^2G(s_8s_0)).
\ee
As soon as in the chiral limit $J_{\mu5}^8$ should be conserved,
from Eq.(\ref{DivJ8}) follows that coefficients of the terms  $m_{\eta'}^2,
m_G^2$ must decrease at least as $(m_\eta/m_{\eta',G})^2$. More specifically, we expect the following limits for the terms of Eq.(\ref{DivJ8}):

\be \label{Constr-1}
\frac{|s_8s_0|}{|s_3c_8+c_3c_0s_8|}\lesssim \left(\frac{m_{\eta}}{m_G}\right)^2.
\ee

\section{Abelian axial anomaly and charmonium decays ratio}\label{Anomaly&R}
In our paper the dispersive form of the anomaly sum rule will be
extensively used, so we remind briefly the main points of this
approach (see e.g. review \cite{Ioffe:2006ww} for details).

Consider a  matrix element of a transition of the axial current
to two photons with momenta $p$ and $p'$ \be
 T_{\mu \alpha \beta} (p, p') = \langle p, p' \vert J_{\mu 5}
\vert 0 \rangle \;. \ee

The general form  of $T_{\mu \alpha \beta}$ for a case $p^2=p'^2$
can be represented in terms of structure functions (form factors):

\begin{multline}  T_{\mu \alpha \beta}(p, p') = F_1(q^2) q_{\mu}
\epsilon_{\alpha \beta \rho \sigma} p_{\rho} p'_{\sigma} +  \\
\frac{1}{2} F_2 (q^2) [\frac{p_{\alpha}}{p^2} \epsilon_{\mu \beta
\rho \sigma}p_{\rho}p'_{\sigma} - \frac{p'_{\beta}}{p^2}
\epsilon_{\mu \alpha \rho
\sigma}p_{\rho}p'_{\sigma}-\epsilon_{\mu \alpha \beta
\sigma}(p-p')_{\sigma}],
\end{multline}
where $q=p+p'$. The functions
$F_1(q^2)$, $F_2(q^2)$ can be described by dispersion relations
with no subtractions and anomaly condition in QCD results in the
sum rule:

\be \label{sumrule} \int\limits^{\infty}_{0}~ Im~F_1(q^2) dq^2 =
2\alpha N_c\sum e_q^2\;, \ee
where $e_q$ are quark electric
charges and $N_c$ is the number of colors. This sum rule \cite{Frishman:1980dq} was
developed by Ji\v{r}\'{\i} Ho\v{r}ej\v{s}\'{\i} \cite{Horejsi:1985qu}, and later generalized \cite{Veretin:1994dn}. Notice that in QCD this equation
does not have any perturbative corrections
\cite{Adler:1969er}, and it is expected that it does not have any
non-perturbative corrections as well due to the 't~Hooft's consistency
principle \cite{Horejsi:1994aj}. It will be important for us that
as $q^2 \to \infty$ the function $ImF_1(q^2)$ decreases as
$1/q^4$ (see discussion in Ref. \cite{Klopot:2009cm}). Note also that the relation (\ref{sumrule}) contains
only mass-independent terms, which is especially important for
the 8th component of the axial current $J_{\mu5}^8$ containing
strange quarks:

\be
 J^{8}_{\mu 5} = \frac{1}{\sqrt{6}}(\bar{u} \gamma_{\mu} \gamma_5 u + \bar{d}
\gamma_{\mu} \gamma_5 d - 2\bar{s} \gamma_{\mu} \gamma_5 s)\;.
\ee

The general sum rule (\ref{sumrule}) takes the form:

\be \label{sumrule8} \int\limits^{\infty}_{0}~ Im~F_1(q^2) dq^2 =
\frac{2}{\sqrt{6}}\alpha (e^2_u + e^2_d - 2e^2_s) N_c =
\sqrt{\frac{2}{3}}\alpha  \;, \ee
 where $e_u=2/3$,\; $e_d=e_s=-1/3$, $N_c=3$.

In order to separate the form factor $F_1(q^2)$, multiply $T_{\mu
\alpha \beta} (p, p')$ by $q_\mu /q^2$. Then, taking the imaginary
part of $F_1(q^2)$, using the expression for $\partial_\mu
J_{\mu5}^8$ from Eq.(\ref{DivJ_FUMPhi}) and unitarity  we get:
\begin{multline}
 Im F_1(q^2)=Im~ q_{\mu}\frac{1}{q^2} \langle 2\gamma \mid J^{(8)}_{\mu 5}\mid0 \rangle = \\
 -\frac{f_8}{q^2}\langle 2\gamma \mid[m_{\eta}^2\eta(c_8c_3-c_0s_3s_8) +
 m_{\eta'}^2\eta'(s_3c_8+c_3c_0s_8) + m_{G}^2 G s_8s_0 ]\mid0 \rangle= \\
 \pi f_8[A_{\eta}\delta(q^2-m_{\eta}^2)(c_8c_3-c_0s_3s_8)+
 A_{\eta'}\delta(q^2-m_{\eta'}^2)(s_3c_8+c_3c_0s_8)+
 A_{G}\delta(q^2-m_{G}^2)(s_8s_0)].
\end{multline}

If we employ the sum rule (\ref{sumrule8}), we obtain a simple
equation:
\be \label{J8_anom_final}
(c_8c_3-c_0s_3s_8)+\beta(s_3c_8+c_3c_0s_8)+\gamma(s_8s_0)=\xi,
\ee
where

\be
\beta\equiv\frac{A_{\eta'}}{A_{\eta}}=\sqrt{\frac{\Gamma_{\eta'\to
2\gamma}}{\Gamma_{\eta \to
2\gamma}}\frac{m_{\eta}^3}{m_{\eta'}^3}}, \;\;\;
\gamma\equiv\frac{A_{G}}{A_{\eta}}=\sqrt{\frac{\Gamma_{G\to
2\gamma}}{\Gamma_{\eta \to 2\gamma}}\frac{m_{\eta}^3}{m_{G}^3}},
\ee \be \xi\equiv\sqrt{\frac{\alpha^2
m^3_{\eta}}{96{\pi}^3\Gamma_{{\eta} \to 2\gamma}}\frac{1}{f^2_8}}
,\;\;\;\;
\Gamma_{\eta\to2\gamma}=\frac{m_{\eta}^3}{64\pi}A_\eta^2\;. \ee

Note that if we include higher resonances in this
equation, they will be suppressed as $1/m^2_{res}$ by
virtue of the mentioned above asymptotic behavior of
$F_1(q^2)\propto 1/q^4$.
For the last two terms in (\ref{J8_anom_final}) we can specify this constraint as follows:
\be \label{Constr-2}
\frac{|s_8s_0|}{|s_3c_8+c_3c_0s_8|}\lesssim \frac{\beta}{\gamma}\left(\frac{m_{\eta'}}{m_G}\right)^2.
\ee

As an additional experimental constraint
we use, following \cite{Akhoury:1987ed,Ball:1995zv}, the
data of the decay ratio $R_{J/\Psi}=(\Gamma(J/\Psi)\to
\eta'\gamma)/(\Gamma(J/\Psi)\to \eta\gamma)$.

As it was pointed out in \cite{Novikov:1979uy}, the radiative
decays $J/\Psi \to \eta (\eta')\gamma$ are dominated by
non-perturbative gluonic matrix elements, and the ratio of the
decay rates $R_{J/\Psi}=(\Gamma(J/\Psi)\to
\eta'\gamma)/(\Gamma(J/\Psi)\to \eta\gamma)$ can be expressed as
follows: \be \label{RJP1} R_{J/\Psi}=\left|\frac{\langle0\mid
G\widetilde{G}\mid\eta'\rangle}{\langle0\mid
G\widetilde{G}\mid\eta\rangle}\right|^2\left(\frac{p_{\eta'}}{p_{\eta}}\right)^3,
\ee where
$p_{\eta(\eta')}=M_{J/\Psi}(1-m^2_{\eta(\eta')}/M^2_{J/\Psi})/2$.
The advantage of this ratio is expected smallness of
perturbative and non-perturbative corrections.

The divergencies of singlet and octet components of the axial current
in terms of quark fields can be written as:

\be \label{RRPP1}
\partial_\mu J_{\mu5}^8=\frac{1}{\sqrt{6}}(m_u \overline{u}\gamma_5u+ m_d \overline{d}\gamma_5d- 2m_s\overline{s}\gamma_5s ),
\ee
\be  \label{RRPP2}
\partial_\mu J_{\mu5}^0=\frac{1}{\sqrt{3}}(m_u \overline{u}\gamma_5u+ m_d \overline{d}\gamma_5d + m_s\overline{s}\gamma_5s ) +
\frac{1}{2\sqrt{3}}\frac{3\alpha_s}{4\pi}G\widetilde{G}.
\ee

Following \cite{Akhoury:1987ed}, neglect the contribution of u-
and d- quark masses, then the matrix elements of the anomaly term
between the vacuum and $\eta, \eta'$ states are:
  \be \label{RJP2}
  \frac{\sqrt{3}\alpha_s}{8\pi}\langle 0\mid G\widetilde{G}\mid\eta \rangle=\langle 0\mid \partial_\mu J^{(0)}_{\mu 5}\mid \eta \rangle+\frac{1}{\sqrt{2}}\langle 0\mid \partial_\mu J^{(8)}_{\mu 5}\mid \eta \rangle,
  \ee
  \be \label{RJP3}
  \frac{\sqrt{3}\alpha_s}{8\pi}\langle 0\mid G\widetilde{G}\mid\eta' \rangle=\langle 0\mid \partial_\mu J^{(0)}_{\mu 5}\mid \eta' \rangle+\frac{1}{\sqrt{2}}\langle 0\mid \partial_\mu J^{(8)}_{\mu 5}\mid \eta' \rangle.
  \ee

Using Eq. (\ref{DivJ_FUMPhi}), (\ref{RJP1}), (\ref{RJP2}), (\ref{RJP3}) we deduce:

\be \label{RJP4Final}
R_{J/\Psi}=\left[\frac{f_0(-s_3s_8+c_3c_8c_0)+\frac{1}{\sqrt{2}}
f_8 (s_3c_8+c_3c_0s_8)}{f_0
(-c_3s_8-c_8c_0s_3)+\frac{1}{\sqrt{2}}f_8(c_8c_3-c_0s_3s_8)}\right]^2  \times \left(\frac{m_{\eta'}}{m_\eta}\right)^4\left(\frac{p_{\eta'}}{p_{\eta}}\right)^3.
\ee

\section{Analysis}\label{Analysis}

For further analysis it is convenient to rewrite the equations (\ref{J8_anom_final}), (\ref{RJP4Final}) in terms of angles $\theta_1\equiv\theta_8+\theta_3$, $\theta_8$ and $\theta_0$:

\be \label{Anom-f}
\frac{1}{2}(c_1+c_2-c_0(c_2-c_1))+\frac{\beta}{2}(s_1-s_2+c_0(s_1+s_2))+ \gamma (s_8s_0)=\xi.
\ee

\begin{equation}
\label{R-f}
R_{J/\Psi}=\left[\frac{f_0(c_1-c_2+c_0(c_1+c_2))+\frac{1}{\sqrt{2}}
f_8 (s_1-s_2+c_0(s_1+s_2))}{f_0
(-s_1-s_2-c_0(s_1-s_2))+\frac{1}{\sqrt{2}}f_8(c_1+c_2-c_0(c_2-c_1))}\right]^2\left(\frac{m_{\eta'}}{m_\eta}\right)^4\left(\frac{p_{\eta'}}{p_{\eta}}\right)^3,
\end{equation}
where $\theta_2\equiv2\theta_8-\theta_1$.

The angles $\theta_1$, $\theta_8$, $\theta_0$ have the explicit physical meaning. From the
definition (\ref{MixMat}) of the  mixing matrix $\mathbf{U}$  one can see that the angle $\theta_1$ describes the overlap in the $\eta-\eta'$ system with an accuracy $\sim \theta_0^2/2$ and
coincides with their mixing angle as $\theta_0 \to 0$. At the same time $\theta_0$ is responsible for the glueball admixture to $\eta-\eta'$ system, and $s_8s_0$  describes the contribution of the glueball state $G$ to the octet component of axial current $\partial J_{\mu5}^8$ only.

In the further analysis we will use the following assumptions:

I) As we discussed in Sec. \ref{Mixing},  the last term in (\ref{DivJ8}) should be suppressed as $(m_{\eta}/m_G)^2$. So we impose the following constraint:

\be \label{Constr-1a}
\frac{|s_8s_0|}{|s_3c_8+c_3c_0s_8|}\lesssim \left(\frac{m_{\eta}}{m_G}\right)^2.
\ee

II) In sec 3 we found another constraint, which follows from the asymptotic behavior of $Im F_1$ (see \ref{Constr-2}):

\be \label{Constr-2a}
\frac{|s_8s_0|}{|s_3c_8+c_3c_0s_8|}\lesssim \frac{\beta}{\gamma}\left(\frac{m_{\eta'}}{m_G}\right)^2.
\ee

III) In our numerical analysis  we suppose that $\gamma$ cannot exceed 1  (i.e.  $\Gamma_{G\to
2\gamma}/m_G^3\lesssim \Gamma_{\eta\to2\gamma}/m_{\eta}^3$).
This restriction corresponds to the assumption  that 2-photon
decay widths of pseudoscalar mesons grow like the third power of
their masses, or in other words,  the glueball coupling to quarks
is of the same order as for the meson octet states.

IV) We accept that the decay constants obey the relation $f_8\gtrsim f_0\gtrsim f_\pi$ (for various kinds of justification see, e.g., \cite{Feldmann:1998sh, Leutwyler:1997yr}).

For the purposes of numerical analysis, the values of $R_{J/\Psi}$ ($R_{J/\Psi}=4.8\pm0.6$), masses and two-photon decay widths of $\eta$, $\eta'$ mesons are taken from PDG \cite{Amsler:2008zzb}. Using the values $m_\eta, m_{\eta'}, \Gamma_{\eta\to2\gamma}, \Gamma_{\eta'\to2\gamma}$, we see that the relation for the constraint (\ref{Constr-1a}) is more strict than the constraint (\ref{Constr-2a}). Supposing the minimal mass of the glueball to be of order $m_G\simeq3 m_\eta \simeq1.5$ GeV, we get the estimation:
\be \label{Constr-f}
 |s_8s_0|/|s_3c_8+c_3c_0s_8|\lesssim 0.1.
\ee

\begin{figure}[t]
  \begin{center}
    \subfigure[$(f_8,f_0)=(1.0,1.0)f_\pi$,$\theta_0=0^\circ$]{\label{figa}\includegraphics[width=0.3\textwidth]{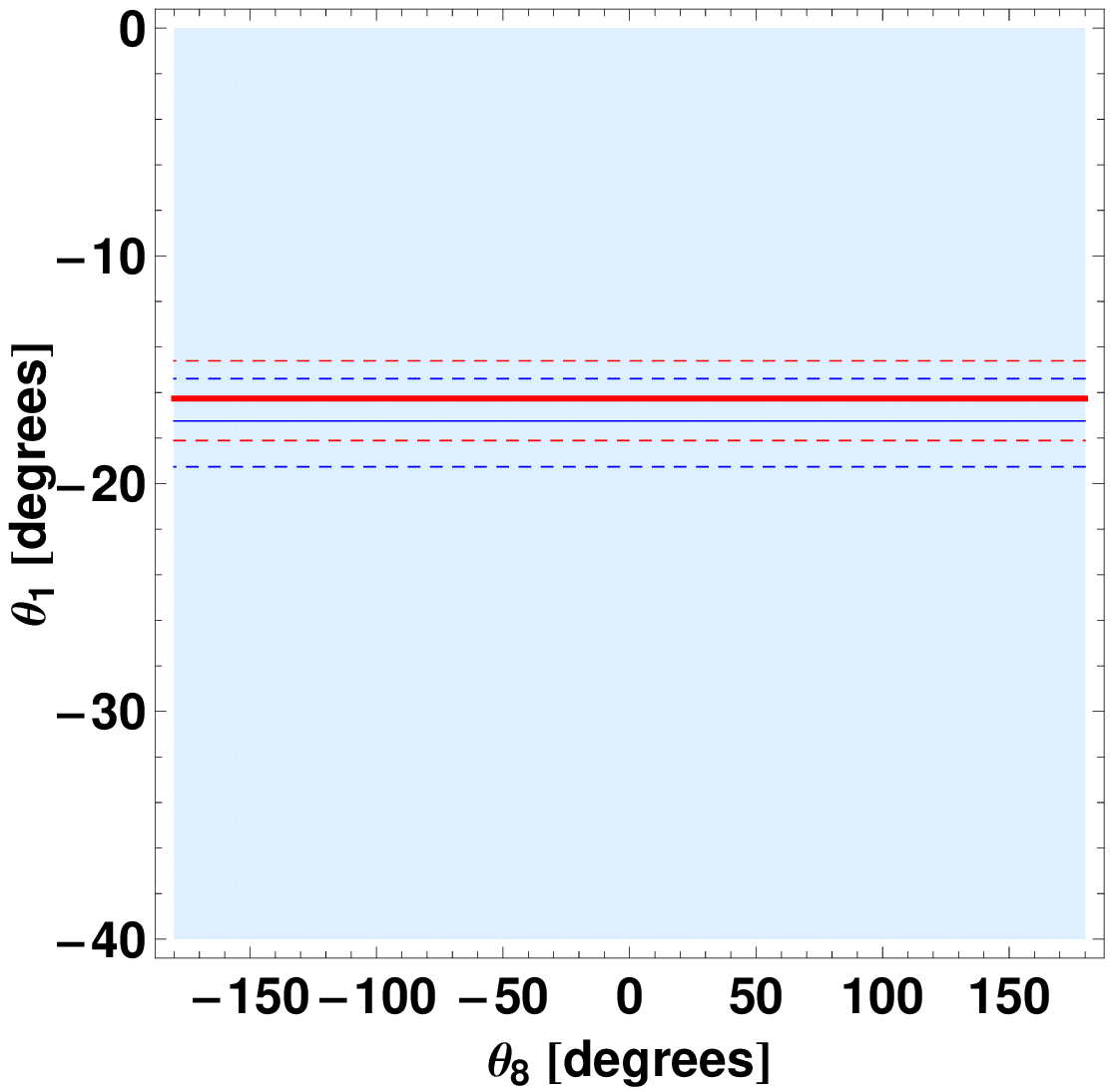}}
    \subfigure[$(f_8,f_0)=(1.0,1.0)f_\pi$,$\theta_0=5^\circ$]{\label{figb}\includegraphics[width=0.3\textwidth]{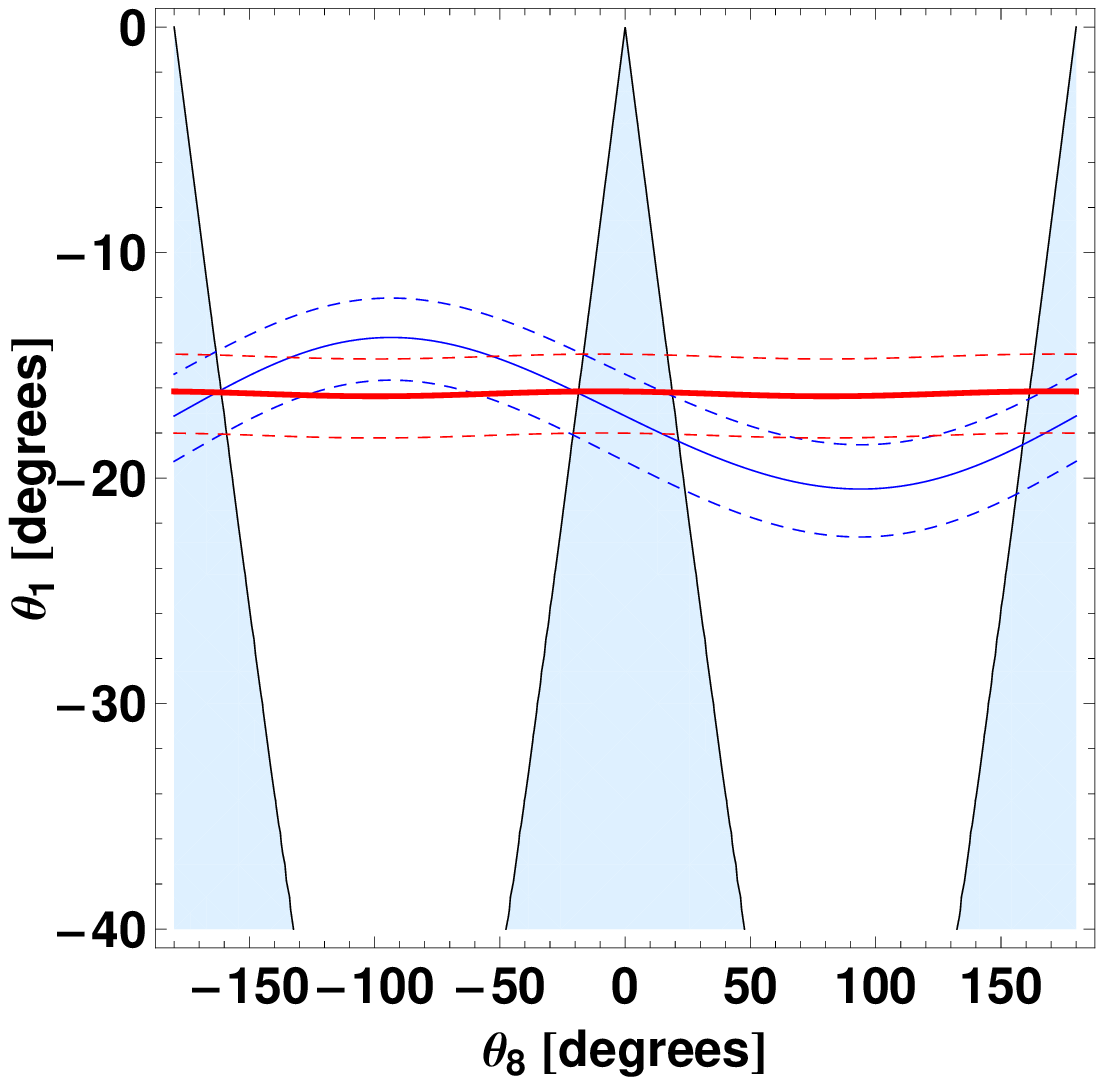}}
    \subfigure[ $(f_8,f_0)=(1.0,1.0)f_\pi$,$\theta_0=30^\circ$]{\label{figc}\includegraphics[width=0.3\textwidth]{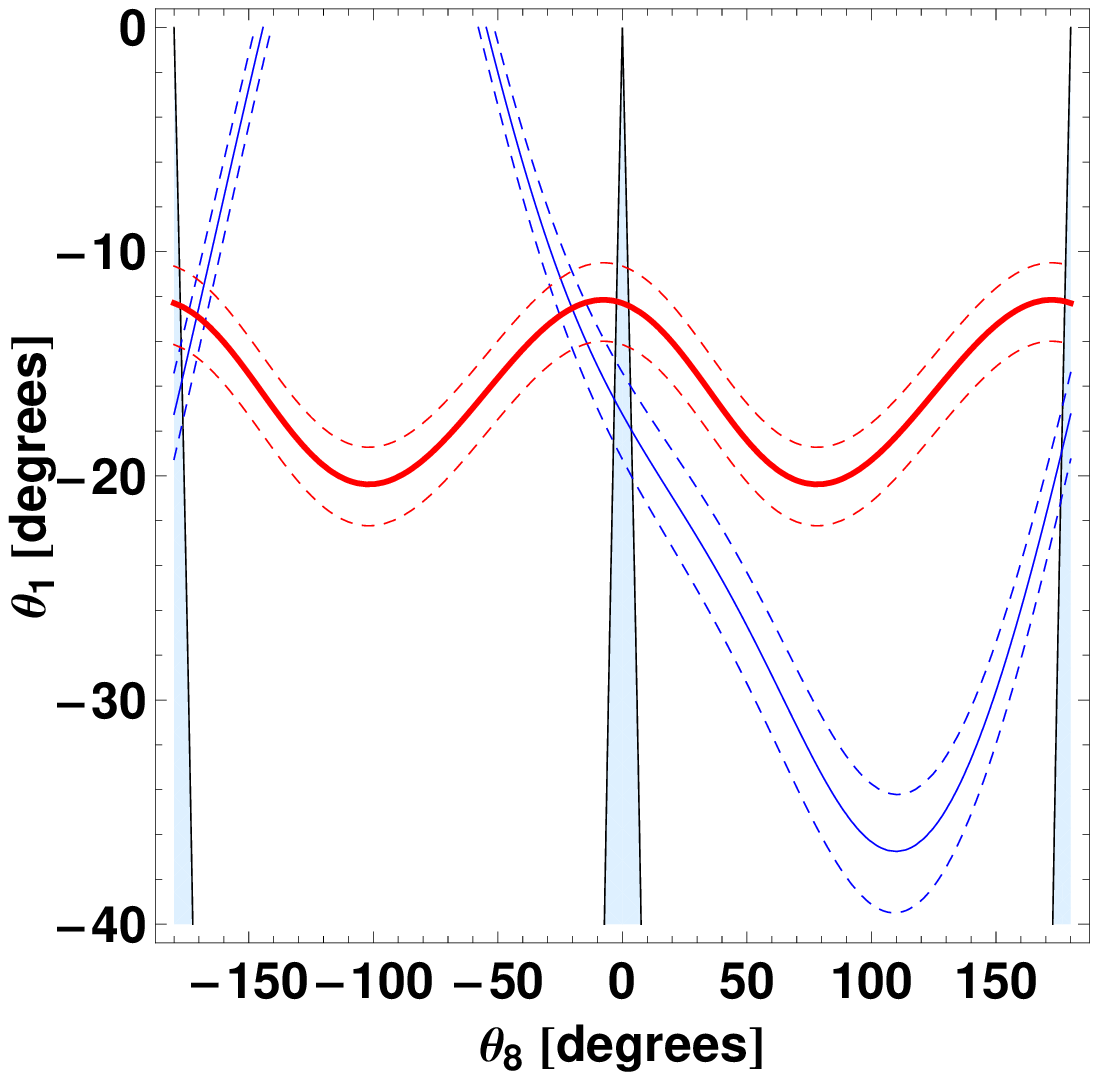}}
    \subfigure[$(f_8,f_0)=(1.1,1.0)f_\pi$,$\theta_0=0^\circ$]{\label{figd}\includegraphics[width=0.3\textwidth]{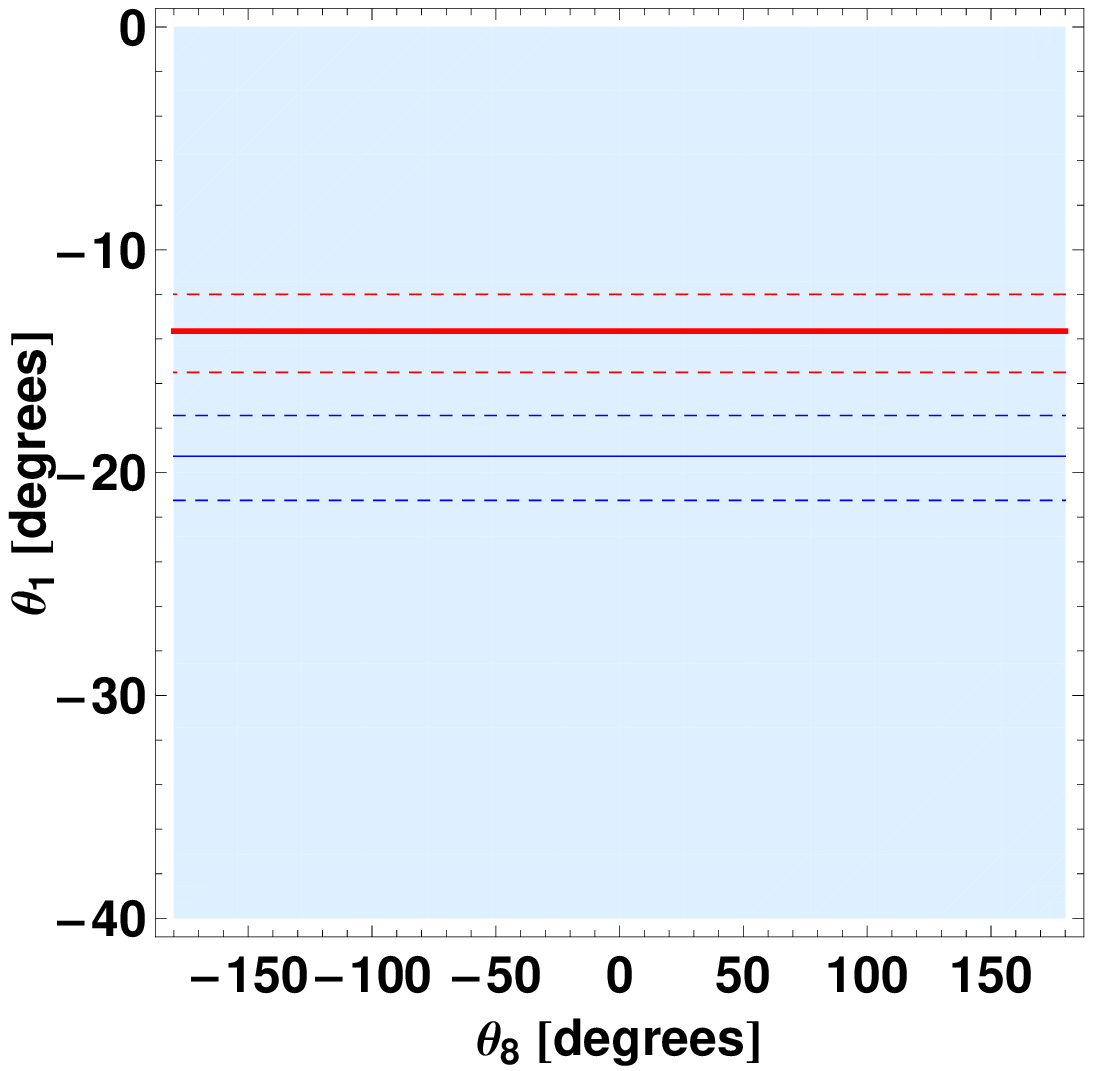}}
    \subfigure[$(f_8,f_0)=(1.1,1.0)f_\pi$,$\theta_0=5^\circ$]{\label{fige}\includegraphics[width=0.3\textwidth]{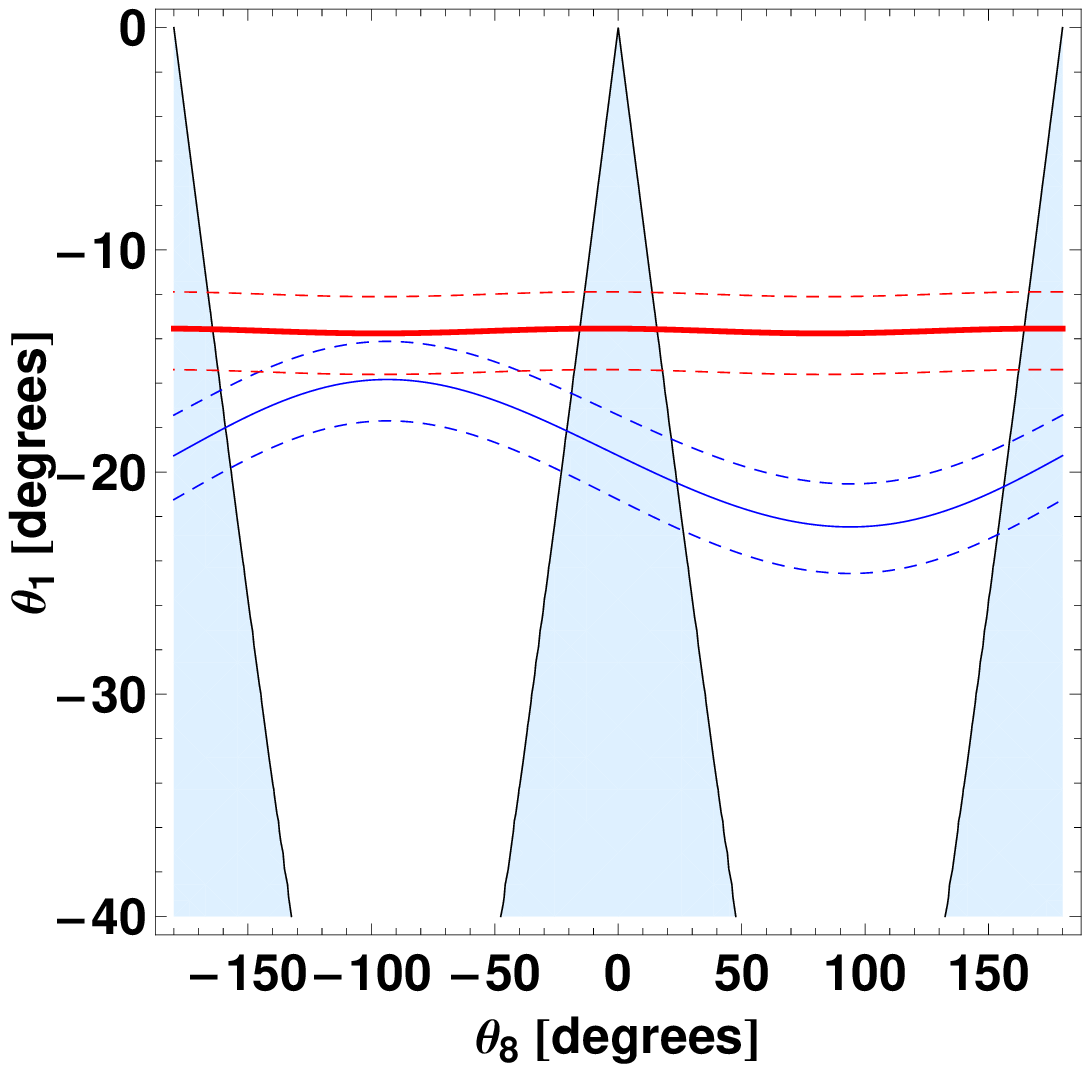}}
    \subfigure[ $(f_8,f_0)=(1.1,1.0)f_\pi$,$\theta_0=30^\circ$]{\label{figf}\includegraphics[width=0.3\textwidth]{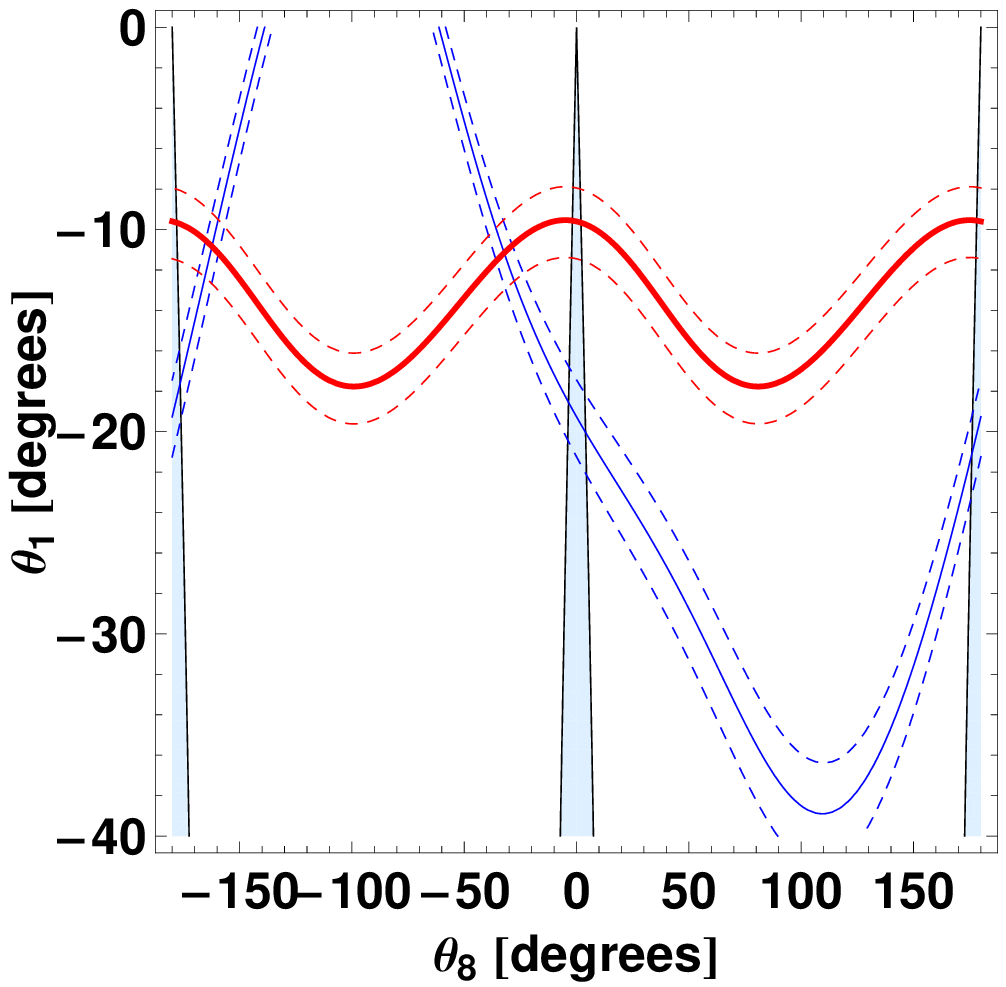}}
  \end{center}
  \caption{The solutions of the Eq. (\ref{Anom-f}) (thin curves, blue online) and (\ref{R-f})(thick curves, red online) with the experimental uncertainties (dashed curves) for different values of the parameters $f_8, f_0$ and $\theta_0$. The shaded area indicates the region, where the relation (\ref{Constr-f}) is valid.}
  \label{fig}
\end{figure}

On Fig. \ref{fig} the plots of the equations (\ref{Anom-f}) and (\ref{R-f})  in the parameter space ($\theta_8, \theta_1$)  are shown for different values of decay constants $f_8$, $f_0$ and mixing angle $\theta_0$. The dashed curves denote experimental uncertainties. The intersection points of the curves represent the solutions of both equations (\ref{Anom-f}),(\ref{R-f}). The filled area indicates the region, where the constraint (\ref{Constr-f}) is valid. The plotted range of angle $\theta_1$ is limited to the physically interesting region, where the solution for relatively small angles $\theta_0$ exists. Let us note for completeness, that there is another solution for $\theta_1 \sim 90^\circ, \theta_0 \gtrsim 50^\circ$ which does not seem to have a physical sense.

The numerical analysis shows, that the solution of the equations (\ref{Anom-f}) and (\ref{R-f}) satisfying the mentioned above constraint is possible only for rather small mixing angle $\theta_0$ and for decay constants $f_8$, $f_0$ close to each other and close to $f_\pi$: for $f_8/f_\pi=f_0/f_\pi=1.0$ the possible range of mixing angle $\theta_0$ is $\theta_0=(0\div25)^\circ$ (see Fig. \ref{figa}-\ref{figc} for demonstration), for $f_8/f_\pi=f_0/f_\pi=1.1$ the possible range of mixing angle $\theta_0$ is $\theta_0=(0\div20)^\circ$.

There is no solutions for decay constant values $f_8/f_\pi=1.1, f_0/f_\pi=1.0$ for any $\theta_0$ (see Fig. \ref{figd}-\ref{figf} for demonstration), and  for any $f_0 \lesssim f_8$ in case of $f_8/f_\pi\geq1.2$.
The obtained results are quite stable: even if we relax the constraint (\ref{Constr-f}) making its r.h.s. several times larger, all the conclusions are preserved.

Note finally, that this result is in contradiction with the prediction for the decay constant $f_8/f_\pi=1.34$  \cite{Kaiser:1998ds} obtained in the Large $N_c$ ChPT.

\section{Conclusion} \label{Conclusion}

In this paper we studied what can be learnt about the mixing in the pseudoscalar sector from the dispersive approach to axial anomaly.

Our analysis shows that the equal values of axial current coupling constants $f_8$ and $f_0$ are favorable which may be considered as a manifestation of $SU(3)$ and chiral symmetry. Moreover, with a less definiteness the relation $f_\pi\approx f_8 \approx f_0$ \cite{Klopot:2009cm} is also supported.

The analysis demands $f_8<1.2 f_\pi$ which deviates at 10\% level from the results of calculations within the chiral perturbation theory ($f_8=1.34 f_\pi$)\cite{Kaiser:1998ds}.

The value of the mixing angle $\theta_0$, which is responsible for the glueball admixture to the $\eta-\eta'$, is limited to $\theta_0<25^\circ$ for $(f_8,f_0)=(1.0,1.0)f_\pi$ and to $\theta_0<20^\circ$ for $(f_8,f_0)=(1.0,1.0)f_\pi$.

The improvement of the experimental data of $R_{J/\Psi}$ can significantly limit the constraints for the parameters $\theta_0$, $\theta_8$ and $f_8,f_0$.

We thank J.~Ho\v{r}ej\v{s}\'{\i}, B.~L.~Ioffe and M.~A.~Ivanov for useful comments and discussions. Y.~K. and O.~T. gratefully acknowledge the  organizers of the workshop for hospitality and support.
This work was supported in part by RFBR (Grants
09-02-00732, 09-02-01149), by the funds from EC to the project
"Study of the Strong Interacting Matter" under contract N0.
R113-CT-2004-506078 and by  CRDF  Project  RUP2-2961-MO-09.



\end{document}